\documentclass[review,11pt]{elsarticlemod}
\usepackage{amsmath,amssymb,amsfonts}
\usepackage{url}
\usepackage{graphicx,tikz,pgf,epsf}
\usepackage[ruled, vlined]{algorithm2e}
\newcommand{\pf}{{\bf Proof: }}
\newtheorem{theorem}{Theorem}
\newtheorem{lemma}{Lemma}
\newtheorem{proposition}{Proposition}

\begin{document}
\begin{frontmatter}
\title{Eavesdropping in Semiquantum Key Distribution Protocol}
\author{Arpita Maitra\fnref{a1}}
\author{Goutam Paul\corref{cor1}\fnref{a2}}
\fntext[a1]{Applied Statistics Unit,
Indian Statistical Institute,
203 B T Road, Kolkata 700 108, India.
Email: arpita76b@rediffmail.com.
{\em The work of this author was supported by the WOS-A fellowship of the 
Department of Science and Technology, Government of India.}}
\cortext[cor1]{Corresponding author.}
\fntext[a2]{Department of Computer Science and Engineering,
Jadavpur University, Kolkata 700 032, India.
Email: goutam.paul@ieee.org.
{\em The work of this author was done in part during his visit at RWTH Aachen, 
Germany as an Alexander von Humboldt Fellow.}}

\begin{abstract}
In semiquantum key-distribution (Boyer et al.) Alice has the same capability
as in BB84 protocol, but Bob can measure and prepare qubits
only in $\{|0\rangle, |1\rangle\}$ basis and reflect any other qubit. We study
an eavesdropping strategy on this scheme that listens to the channel in both
the directions. With the same level of disturbance induced in the channel, Eve
can extract more information using our two-way strategy than what can be
obtained by the direct application of one-way eavesdropping in BB84.
\end{abstract}

\end{frontmatter}

\section{Introduction}
The BB84 protocol~\cite{bb84} is used by Alice and Bob to settle on a secret 
classical bit-string over an insecure quantum channel where Eve can have 
access. There are a number of important papers that 
analyze this scheme and we refer to~\cite{biham97,cnot1,bb284,gisin,fuchs97} 
and the references therein for further reading. The BB84 protocol~\cite{bb84} 
uses the bases $Z = \{|0\rangle, |1\rangle\}$ and 
$X = \{|+\rangle, |-\rangle\}$, where
$|+\rangle = \frac{|0 \rangle + |1\rangle}{\sqrt{2}}$ and
$|-\rangle = \frac{|0 \rangle - |1\rangle}{\sqrt{2}}$. In this case, Bob has 
the capability of measuring the qubits in either $Z$ or $X$ basis.

In~\cite{semi1}, it has been considered that Bob has limited capability
and he can do the following.

(1) Whenever a qubit passes through Bob, he can let it go undisturbed,
or in other words he can reflect the qubit to Alice (CTRL bits).

(2) Otherwise he can measure the qubit in the $Z$ basis and
prepare a fresh qubit in the same basis and send it to Alice
(SIFT bits).

\noindent Based on this limited capability of Bob, semiquantum key distribution 
has been presented in~\cite{semi1} and later it has been analyzed in more 
detail in~\cite{semi2}. The authors call the $Z$ basis as the classical basis,
as it has one-to-one correspondence with the classical bits. Bob
is called {\em classical} since he prepares and measures qubits in this 
basis only. Alice is not classical, as she needs to deal with
quantum superposition of the computational basis states. Thus the protocol
is called {\em semiquantum}. The exact protocol~\cite{semi1} is described
in Algorithm~\ref{algo1}.

\incmargin{1em}
\restylealgo{boxed}
\begin{algorithm}[htbp]
\SetLine
\BlankLine
{\scriptsize
\nl Alice generates $N = 8n(1 + \delta)$ many qubits randomly in $Z$ basis or 
$X$ basis\;
\nl For each qubit received at Bob's end, he chooses randomly
either to reflect it (CTRL) or to measure it in $Z$ basis and resend it in 
the same state he measured (SIFT)\;
\nl Alice measures each qubit in the basis she sent\;
\nl Alice publishes which are the $Z$ bits she sent and Bob publishes which
ones he chose to SIFT\;
\nl Alice checks the error-rate in the CTRL bits and aborts the protocol if
the error-rate in either of the basis is more than some predefined threshold
value\;
\nl Alice randomly chooses $n$ SIFT bits as TEST bits and publishes them\;
\nl Bob publishes the values of these TEST bits\;
\nl Alice checks the error rate in these bits and aborts if the error-rate is 
more than some predefined value\;
\nl Alice and Bob select the first $n$ remaining SIFT bits to be used
as INFO bits\;
\nl Alice publishes ECC and PA data and then she and Bob use them to
extract the final $m$-bit secret key from $n$-bit INFO string\;
}
\caption{Semiquantum Key Distribution with Classical Bob~\cite{semi1}.}
\label{algo1}
\end{algorithm}
\decmargin{1em}

In~\cite{semi1,semi2}, the authors proved that the protocol is robust. This is 
in the sense that for any attack inducing no error on TEST bits (a subset of 
the SIFT bits used for error correction~\cite{semi1}) and CTRL bits,
Eve's final state is independent of the qubits chosen by Bob as SIFT and the 
states sent by Alice. However, they did not give any quantitative estimate 
connecting the disturbance experienced by Alice and the information leakage at 
Eve's end. In~\cite[Section I]{semi2}, the authors made the following comment:
\begin{quote}
{\sf Note that our result does not imply that Eve cannot gain a large amount 
of information by inducing a very small ͑but nonzero͒ noise on qubits. 
It is however our belief that such a discontinuity of
``information versus disturbance" does not occur, but the
question is beyond the scope of this paper and is left for
future research.}
\end{quote}
This gives us motivation to investigate the exact relationship between
the disturbance and information leakage under certain eavesdropping model.
In the semiquantum protocol~\cite{semi1}, the qubits need to travel in two 
directions and thus the eavesdropper has the advantage to look into the qubits 
twice instead of once as in the case of BB84~\cite{bb84}. 

In this regard, let us refer to a comment~\cite{comment1} on the semiquantum 
protocol~\cite{semi1} and the corresponding response~\cite{response1}
that are related to actual implementation issues. 
In~\cite{comment1}, it has been pointed out that if the qubits are implemented
by photons with some wavelength $\lambda$ say, then the eavesdropper can 
cleverly modify the wavelength of the photons (flying from Alice to Bob) 
from $\lambda$ to $\lambda + \delta\lambda$ without changing the polarization. 
For each SIFT photon, as Bob measures and recreates a fresh copy, the 
wavelength again becomes $\lambda$ during the return path from Bob to Alice,
while for each CTRL photon, as it is only reflected by Bob, and the wavelength
remains $\lambda + \delta\lambda$. This is exactly the situation, where
Eve can successfully tag the SIFT and CTRL photons separately before the 
public announcement and thus obtains all the secret key (INFO) bits without 
creating any disturbance in a similar
line to the {\em mock protocol} described in~\cite{semi1} itself. 
In the response~\cite{response1} to this comment~\cite{comment1}, the authors
acknowledge this attack based on implementation issues. However, it is 
also pointed in~\cite{response1} that there may be several such attacks on 
various implementations of any Quantum Key Distribution protocol. Naturally,
there would be several countermeasures, e.g., in this case Bob may place a
suitable filter to thwart this attack. To be precise, the attack 
of~\cite{comment1} considers an implementation weakness, while the attack
does not work on theoretical model using perfect qubits, namely, a two-
dimensional Hilbert space~\cite{response1}.

However, the eavesdropping strategy that we present here is not dependent 
on any implementation issue and should work on any perfect implementation where 
the standard model of eavesdropping can be implemented too. Given that, we show
that our two-way eavesdropping strategy extracts more information about the
secret key bits against~\cite{semi1} than that is possible against traditional
BB84~\cite{bb84} for certain range of disturbance. That is, our eavesdropping
strategy works even in idealistic scenario, while the attack of~\cite{comment1} 
works only in a specific realistic scenario only that can also be thwarted
with proper countermeasure.

\subsection{Eavesdropping Model}
\label{bg}
In this paper, we consider the same symmetric incoherent optimal eavesdropping 
model of~\cite{fuchs97} that was used for the traditional BB84 
protocol~\cite{bb84}. It is symmetric, because there will be equal error 
probability at Bob's end corresponding to different bases and it is
incoherent as Eve works with each individual qubit. 

In general, Alice sends a qubit $|\mu\rangle$ to Bob and Eve lets a 
four dimensional probe $|W\rangle$ of two qubits (as 
in~\cite[Section III]{fuchs97}) that interacts unitarily with $|\mu\rangle$. 
Eve's measurement is delayed till Alice announces the basis that has been used
(i.e., by that time Bob has already measured the state). 
We can model it as $U(|\mu\rangle, |W\rangle) = |\tau\rangle$, where $U$ is 
the unitary operator and after its application, $|\tau\rangle$ is the entangled
state of the qubit that Alice sent to Bob and the probe applied by Eve. 
Let $D$ be the disturbance in the channel due to the interaction by Eve
and $F=1-D$ be the fidelity. Without loss of generality, one can write the 
eavesdropping interaction for the Z basis as
\begin{equation*}
U(|0\rangle, |W\rangle) =
\sqrt{F}|E_{00}\rangle|0\rangle + \sqrt{D}|E_{01}\rangle|1\rangle,
U(|1\rangle, |W\rangle) =
\sqrt{D}|E_{10}\rangle|0\rangle + \sqrt{F}|E_{11}\rangle|1\rangle.
\end{equation*}
Following the analysis in~\cite{fuchs97,gisin}, it can be
shown that with the optimal eavesdropping strategy, Eve's average success 
probability in correctly guessing a secret bit is
$P_E^{(1)}(D) = \frac{1}{2} + \sqrt{D(1-D)}.$ 
Since the {\em advantage} of the eavesdropper can be defined as the amount by 
which the success probability exceeds the probability of random guessing 
(which, in this case, is $\frac{1}{2}$), the advantage is given by
$A_E^{(1)}(D) = \sqrt{D(1-D)}$.

Note that the strategy of~\cite{fuchs97} can directly be used towards
eavesdropping against the semiquantum protocol~\cite{semi1}. 
However, we like to explore beyond this trivial application of the 
eavesdropping strategy of~\cite{fuchs97} and
consider that the eavesdropper will try to extract information during the 
transmission of qubits both from Alice to Bob and Bob to Alice. 

When Alice sends a qubit $|\mu\rangle$ to Bob, Eve lets a 
probe $|W\rangle$ that interacts unitarily with $|\mu\rangle$. Thus the 
interaction can be modelled as $U(|\mu\rangle, |W\rangle) = |\tau\rangle$.
If Bob performs a measurement, then the 3-qubit entangled state $|\tau\rangle$
collapses to a 3-qubit post-measurement product state $|\tau_m\rangle$.
When the corresponding qubit (either reflected or measured and resent) returns 
from Bob to Alice,  then again Eve tries to interact with 
a probe $|W'\rangle$ and the unitary operation can be written as 
$U'(|\tau''\rangle, |W'\rangle) = |\tau'\rangle$, where $|\tau''\rangle$ is
$|\tau\rangle$ or $|\tau_{m}\rangle$, according as Bob reflects or measures
(respectively), and $|\tau'\rangle$ is a 5-qubit state. 

\section{Analysis of the Binary Symmetric Channels}
One may refer to~\cite{fuchs97} to note that the analysis can be done 
considering the model of Binary Symmetric Channel (BSC). 
For the semiquantum case, we need to consider two cascaded channels, 
the first for the qubits moving from Alice to Bob and the second for
the qubits moving from Bob to Alice, each with error probability
$p$ due to Eavesdropping. 
(There can be eavesdropping inducing different error 
probabilities $p_1, p_2$ in the two channels, and that can be taken care of 
in a similar manner.) Following Section~\ref{bg}, we can write for the 
channel between Alice and Bob
\begin{eqnarray}
\label{u01p}
U(|0\rangle, |W\rangle) & = & 
\sqrt{1-p}|E_{00}\rangle|0\rangle + \sqrt{p}|E_{01}\rangle|1\rangle, \nonumber\\
U(|1\rangle, |W\rangle) & = &
\sqrt{p}|E_{10}\rangle|0\rangle + \sqrt{1-p}|E_{11}\rangle|1\rangle,
\end{eqnarray}
where $W$ is the initial state of the pair of qubits at Eve's hand.

We analyze the SIFT bits and the CTRL bits separately.
\begin{lemma}
\label{sift} 
For the SIFT bits, the round trip channel from Alice to Bob and back to Alice 
is equivalent to a binary symmetric channel with error probability $2p(1-p)$.
\end{lemma}
\pf Refer to Figure~\ref{bsc}. The first BSC corresponds to the transmission 
from Alice to Bob and the second one corresponds to the transmission from Bob 
to Alice.
{\small
\begin{figure}[htb]
\begin{center}
\begin{tikzpicture}[scale=1]

\node at (1.5,0) [below] {$1-p$};
\draw [->] (0,0)  node [left] {1} -- (2.8,0) node [right] {1};
\node at (1.5,2) [above] {$1-p$};
\draw [->] (0,2)  node [left] {0} -- (2.8,2) node [right] {0};

\draw [->] (0,0.2) -- (2.8,1.8);
\draw [->] (0,1.8) -- (2.8,0.2);

\node at (1.7,1.25) [above] {$p$};
\node at (1.7,0.75) [below] {$p$};

\node at (4.5,0) [below] {$1-p$};
\draw [->] (3.2,0) -- (6,0) node [right] {1};
\node at (4.5,2) [above] {$1-p$};
\draw [->] (3.2,2) -- (6,2) node [right] {0};

\draw [->] (3.2,0.2) -- (6,1.8);
\draw [->] (3.2,1.8) -- (6,0.2);

\node at (4.9,1.25) [above] {$p$};
\node at (4.9,0.75) [below] {$p$};

\end{tikzpicture}

\end{center}
\caption{Cascaded BSC model for the SIFT bits}
\label{bsc}
\end{figure}
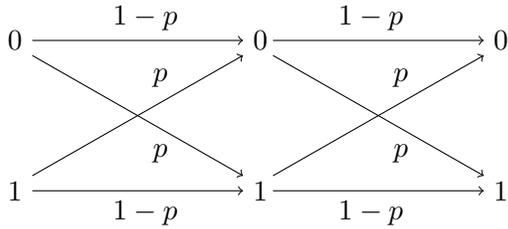
}
Consider one round-trip communication from Alice to Bob and then back 
to Alice. 
If Alice sends 0, the error-path is either $0 \rightarrow 0 \rightarrow 1$
or $0 \rightarrow 1 \rightarrow 1$. The path is symmetric when Alice sends 1.
Thus, the error-probability would be given by $(1-p)p + p(1-p) = 2p(1-p)$. \qed

\begin{lemma}
\label{ctrl}
For the CTRL bits, the round trip channel from Alice to Bob to Alice is
equivalent to a binary symmetric channel with error probability $2p(1-p)$.
\end{lemma}
\pf Without loss of generality, let us consider when Alice sends $0$. The case
when Alice sends $1$ would be symmetric.

According to Equation~\eqref{u01p}, Bob will receive a qubit entangled with
the two bits of Eve, having the three-qubit entangled state
$\sqrt{1-p}|E_{00}\rangle|0\rangle + \sqrt{p}|E_{01}\rangle|1\rangle$.
Bob will send the entangled qubit received as it is to Alice.
Let $W'$ be the initial state of the new pair of qubits at Eve’s
hand with which she will interact unitarily
with the qubit sent from Bob to Alice. This interaction can be written as

$I \otimes U (\sqrt{1-p}|E_{00}\rangle|0\rangle + \sqrt{p}|E_{01}\rangle|1\rangle, |W'\rangle)$

$=\Big((1-p)|E_{00}\rangle|E'_{00}\rangle 
+ p|E_{01}\rangle|E'_{10}\rangle\Big)|0\rangle 
+ \sqrt{p(1-p)}\Big(|E_{00}\rangle|E'_{01}\rangle 
+ |E_{01}\rangle|E'_{11}\rangle\Big)|1\rangle$,

where $E'_{ij}$'s are the new two-qubit probes at Eve's
hand corresponding to Bob sending bit $i$ and Alice receiving bit $j$.
Thus, the probability that Alice measures 0 is $(1-p)^2 + p^2$ and that she
measures 1 is $2p(1-p)$.  \qed

Note that for the SIFT case, one qubit moves from Alice to Bob and a 
corresponding different one moves from Bob to Alice back. In the CTRL case,
it is the same qubit that is travelling in both the directions (Alice to Bob 
and Bob to Alice). It is important to note that the error in the channel 
observed by Alice is the same in both the cases.
This is the reason, the eavesdropping model of~\cite{fuchs97} remains symmetric 
even when it is applied to the semiquantum protocol in both the directions.

Combining Lemma~\ref{sift} and Lemma~\ref{ctrl}, we can write the following 
Theorem.
\begin{theorem}
\label{th11}
The round trip channels from Alice to Bob to Alice 
when Bob measures and sends a fresh qubit (SIFT) and when he just reflects
the received qubit (CTRL) are equivalent, and both act as a binary symmetric
channel with error probability $2p(1-p)$.
\end{theorem}

Let $D_{\mbox{\em one-way}}$ and $D_{\mbox{\em two-way}}$ be the disturbances 
in the one-way BB84 protocol and the two-way semiquantum protocol 
respectively. We take $D_{\mbox{\em one-way}} = D$ and we have shown that 
$D_{\mbox{\em two-way}} = 2p(1-p)$. Both the attacks should be compared in the 
same footing, i.e., Eve's advantages have to be compared at the same 
disturbance values. For this reason, we take
$D_{\mbox{\em two-way}} = D_{\mbox{\em one-way}}$, i.e., $D = 2p(1-p)$.

\section{Detailed Analysis of Two-way Eavesdropping}
Eve will have two guesses for the forward and backward communication.
As discussed in Section~\ref{bg}, for each guess, Eve has the 
success probability $p_E = \frac{1}{2} + \sqrt{p(1-p)} = 
\frac{1}{2} + \epsilon$, where $\epsilon = \sqrt{p(1-p)}$.
If both the guesses give the same outcome, then the probability that the bit
guessed is correct increases. 
Suppose during the forward communication from Alice to Bob, Eve has a success
probability of $\frac{1}{2} + \epsilon_1$ and during the backward
communication from Bob to Alice, Eve has a success probability of
$\frac{1}{2} + \epsilon_2$. 
For a particular bit, let $P_E^{(2,match)}$ be Eve's posterior
probability that the bit sent was $b$, when both her
forward and the backward guesses give the same outcome 
$b \in \{0,1\}$.
The following result is easy to show.
\begin{proposition}
\label{prop}
$P_E^{(2,match)}=\frac{1}{2} + \frac{\epsilon_1+\epsilon_2}{1+4\epsilon_1\epsilon_2}$.
\end{proposition}
Since $\epsilon_i \leq \frac{1}{2}$, we have
$P_E^{(2,match)} \geq \frac{1}{2} + \epsilon_i$, for $i = 1,2$.
This implies that Eve's success probability increases, when she observes 
the same outcome $b$ in both directions and guesses that indeed $b$ was sent.

Let us now substitute $\epsilon_1 = \epsilon_2 = \sqrt{p(1-p)}$, 
where $2p(1-p) = D$.
When Eve observes the same outcome in both the directions, let her
success probability as a function of the disturbance be denoted by
$P_E^{(2,match)}(D)$. Thus, we have the following result.
\begin{lemma}
\label{match}
$P_E^{(2,match)}(D) = \frac{1}{2} + \frac{\sqrt{D/2}}{\frac{1}{2} + D}$.
\end{lemma}
The corresponding advantage is given by
$A_E^{(2,match)}(D) = \frac{\sqrt{D/2}}{\frac{1}{2} + D}$.
One may easily check that 
$\frac{\sqrt{D/2}}{\frac{1}{2} + D} \geq \sqrt{D(1-D)}$ for 
$D \in [0, \frac{1}{2}]$ and in this case, Eve's advantage
for the semiquantum protocol is better than what is achieved in the
eavesdropping~\cite{fuchs97} in BB84 protocol.

When both the cases do not have the same outcome, then the situation is not 
encouraging. In this case one has to consider one of the two outcomes as the
correct guess. Without loss of generality, we accept the outcome of the first 
observation as the correct guess. In this case, 
\begin{equation}
\label{mismatch}
P_E^{(2,mismatch)}(D) = \frac{1}{2} + \sqrt{p(1-p)} = \frac{1}{2} + \sqrt{D/2}.
\end{equation}
It is immediate to note that  
$\sqrt{D/2} \leq \sqrt{D(1-D)}$ for $D \in [0, \frac{1}{2}]$
and in this case Eve suffers with the decreased advantage.

Thus while calculating the average advantage, we consider the following 
strategy: ``{\em if the outcomes observed by Eve in both the directions are 
the same bit $b$, she guesses $b$, else she discards her guess in the backward 
direction and considers only the guess during the forward direction}."
Let $P_E^{(2,avg)}(D)$ denote Eve's average success probability when she
follows the above strategy.
\begin{theorem}
\label{avg}
$P_E^{(2,avg)}(D) = \frac{1}{2} + \frac{\sqrt{D/2}(3+2D)}{2\left(1+2D\right)}$.
\end{theorem}
\pf
It is clear that both the match and the mismatch
happens with probability $\frac{1}{2}$.
Hence the average success probability of Eve is given by
$P_E^{(2,avg)}(D) = \frac{1}{2} P_E^{(2,match)}(D) + \frac{1}{2} P_E^{(2,mismatch)}(D)$.
Substituting values of the probabilities from Lemma~\ref{match} and
Equation~\eqref{mismatch}, we get the result.  \qed

\begin{figure*}[htbp]
\centering
\begin{tabular}{cc}
\includegraphics[width=0.49\textwidth]{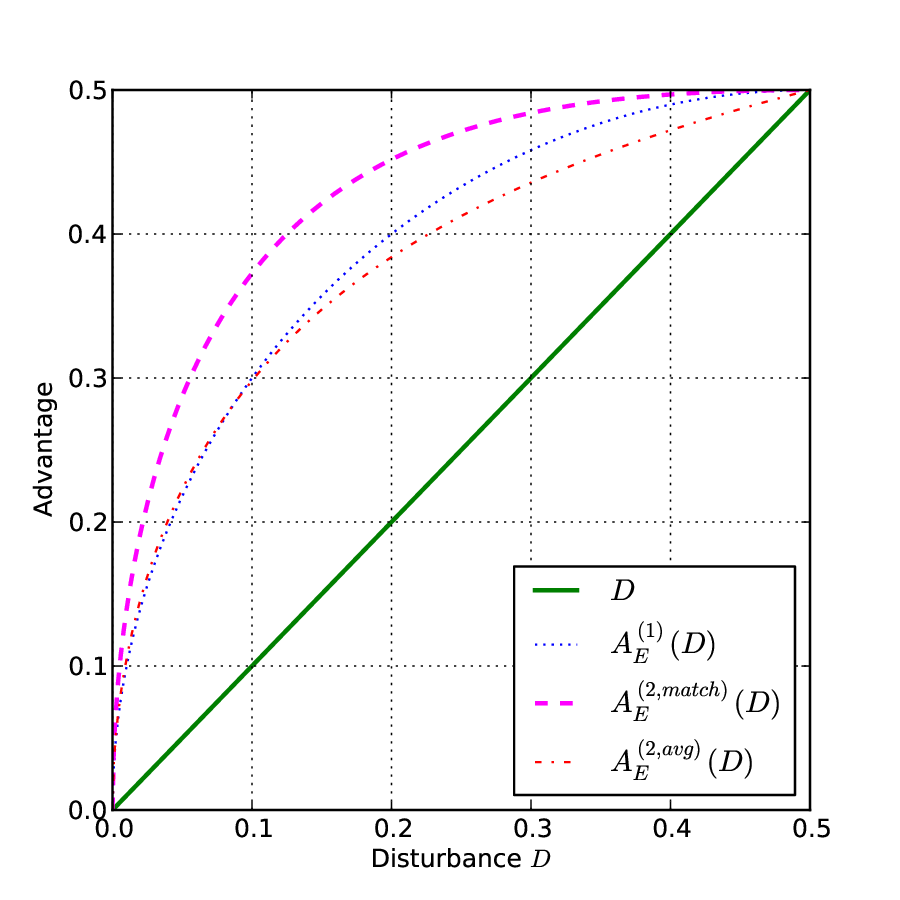}
&
\includegraphics[width=0.49\textwidth]{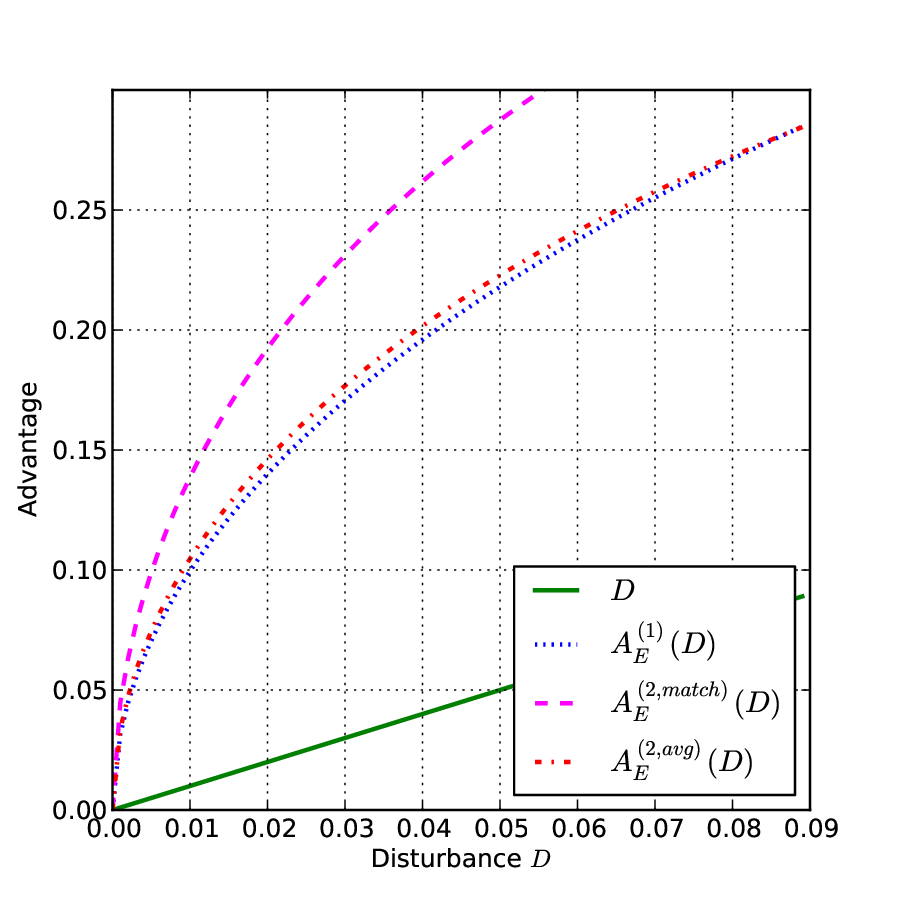}
\end{tabular}
\caption{Advantage of the eavesdropper as a function of disturbance $D$ under different attack models (magnified portion for $0 \leq D \leq 0.09$ is shown on the right).}
\label{advplot}
\end{figure*}

Hence, the average advantage is given by 
$A_E^{(2,avg)}(D) = \frac{\sqrt{D/2}(3+2D)}{2\left(1+2D\right)}$.
In Figure~\ref{advplot}, we plot the advantages for different attack 
strategies versus $D$. Note that $P_E^{(2,avg)}(D) > P_E^{(1)}(D)$, i.e.,
Eve has more advantage in the semiquantum protocol, if $D < 0.0877$ (up to the 
fourth decimal place). This region is shown magnified in the right portion of 
Fig.~\ref{advplot} for a clearer pictorial exposition.

The summarized strategy of eavesdropping on the semiquantum protocol
would be as follows. 

{\em
1. For $D \geq 0.0877$, apply unitary interaction for eavesdropping only 
during the communication between Alice and Bob (same as BB84).

2.  For $D < 0.0877$, apply unitary interaction during both the communications
(Alice to Bob and Bob to Alice) with disturbance in each channel $p$, where
$D = 2p(1-p)$.
If the bits obtained through the two different interactions are 
same, then accept that as the guessed bit.
If the bits obtained through the two different interactions are 
different, then accept the bit obtained during the communication from Alice
to Bob as the guessed one.}

This provides more information to the eavesdropper in certain range in
the semiquantum protocol~\cite{semi1} than in BB84~\cite{bb84} and thus our
two-way eavesdropping on the the semiquantum protocol recovers more information
than what can be obtained using the idea of~\cite{fuchs97} directly as it was 
applied against BB84.

\section{Conclusion}
\label{conclu}
Boyer et al.~\cite{semi1,semi2} introduced the quantum key distribution 
protocol with classical Bob and showed its robustness, but left open any 
analysis regarding how the amount of information leakage to the eavesdropper 
is related to the disturbance caused by her. We analyzed an eavesdropping
strategy on this scheme and explicitly derived eavesdropper's advantage as a 
function of the disturbance. Here our investigation exploits the model 
of~\cite{fuchs97} in both the directions of communication (Alice to Bob and 
Bob to Alice). Our two-way eavesdropping strategy against the semiquantum 
protocol extracts more information on the secret bits than that could be
obtained by direct one-way application of the strategy in~\cite{fuchs97}
that worked on BB84. Other existing eavesdropping strategies on 
BB84~\cite{bb84} may be explored in a similar manner on the semiquantum 
protocol~\cite{semi1,semi2}.

\end{document}